\documentclass[a4paper, oneside, twocolumn, notitlepage, 10pt]{extarticle_ecoc}
\usepackage{ecoc}
\usepackage{comment}
\usepackage{ulem}
\usepackage{stfloats}
\usepackage{booktabs}
\usepackage[disable]{todonotes}
\usepackage[inkscapelatex=false]{svg}

\usepackage{balance}

\begin{document}
\selectlanguage{english}    


\title{Enabling 3GPP-Driven Services Over Optical Transport Network}

\author{
    Renato Ambrosone\textsuperscript{(1)},
    Ramon Casellas\textsuperscript{(2)},
    Alessio Giorgetti\textsuperscript{(3)},
    Riccardo Schips\textsuperscript{(1)},
    Stefano Straullu\textsuperscript{(4)},\\
    Pablo Armingol Robles\textsuperscript{(5)},
    \'Oscar Gonz\'alez de Dios\textsuperscript{(5)},
    Francisco Javier Moreno \textsuperscript{(6)},
    Enrique\\ Fernandez \textsuperscript{(6)},
    Jos\'e Manuel Rivas-Moscoso\textsuperscript{(5)},
    Francesco Aquilino\textsuperscript{(4)},
    Antonino Nespola\textsuperscript{(4)},
    Vittorio Curri\textsuperscript{(1)}
}
\maketitle                  


\begin{strip}
 \begin{author_descr}
 
   \textsuperscript{(1)}Politecnico di Torino, Italy,
   \textcolor{blue}{\uline{renato.ambrosone@polito.it}};
   
   \textsuperscript{(2)}Centre Tecnol\'ogic de Telecomunicacions de Catalunya (CTTC/CERCA), Castelldefels, Spain;\\
    \textsuperscript{(3)}University of Pisa, Pisa, Italy;
    \textsuperscript{(4)}LINKS Foundation, Italy; \\
   \textsuperscript{(5)}Telef\'onica CTIO, Madrid, Spain;
   \textsuperscript{(6)}E-lighthouse  Network Solutions, Cartagena, Spain;
 \end{author_descr}
 \vspace{-2mm}
\end{strip}

\setstretch{1.1}
\renewcommand\footnotemark{}
\renewcommand\footnoterule{}


\begin{strip}
  \begin{ecoc_abstract}
    We demonstrate a multi-layer slicing architecture enabling deployment of 3GPP network slices as self-contained, connected functions atop optical transport slices. The system provides Optical Spectrum as a Service (OSaaS) by integrating disaggregated optical control, multi-tenant orchestration, and telemetry-based optimization via standard interfaces.
    \textcopyright2025 The Author(s)
  \end{ecoc_abstract}

\end{strip}
\section{Introduction}
The increasing demand for dynamic, flexible, and tenant-aware connectivity services in optical networks has accelerated the evolution toward disaggregated, software-controlled architectures~\cite{berthold2008optical}.
Traditional monolithic approaches are being replaced by more modular and open paradigms~\cite{giorgetti2020control}, where heterogeneous devices from multiple vendors can be integrated through standardized interfaces and software-defined control planes. This transformation is particularly relevant in the context of Optical Spectrum-as-a-Service (OSaaS)~\cite{kaeval2024operation}, which aims to virtualize and slice the underlying optical infrastructure to serve diverse tenants with isolated and customizable network views \cite{Casellas2019EnablingNS, Casellas:20}.
Recent initiatives have focused on leveraging on standard APIs such as TAPI~\cite{manso2021tapi}, OpenROADM or OpenConfig~\cite{sgambelluri2020openconfig}, and integrated IP-over-WDM controllers to bridge the gap between optical and packet domains~\cite{de2022mantra}.

\begin{figure*}[b!]
    \centering    \includegraphics[width=1.01\textwidth]{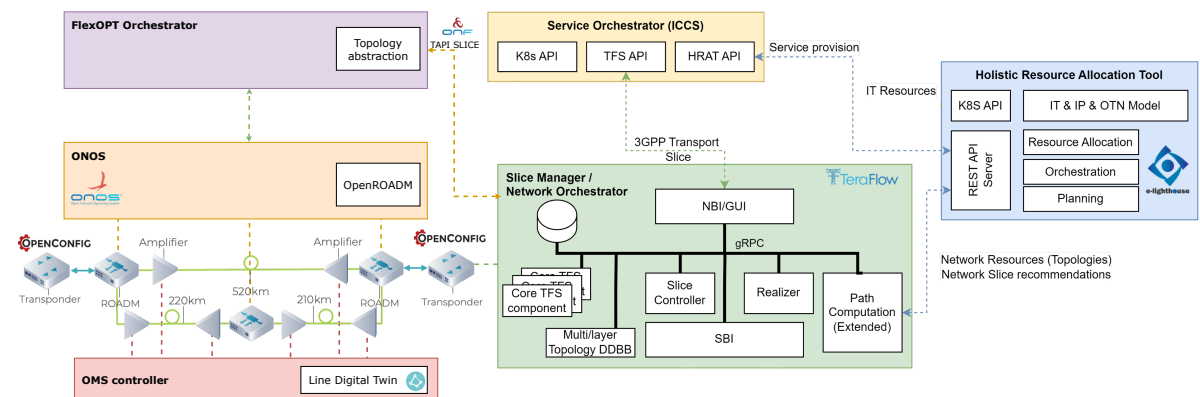}
    \caption{Proposed architecture supporting OSaaS, enabling dynamic and tenant-specific slicing of optical infrastructure with integrated multi-domain and multi-layer orchestration.}
    \label{fig:topology-control}
\end{figure*}
\begin{figure*}[b!]
    \centering    \includegraphics[width=0.98\textwidth]{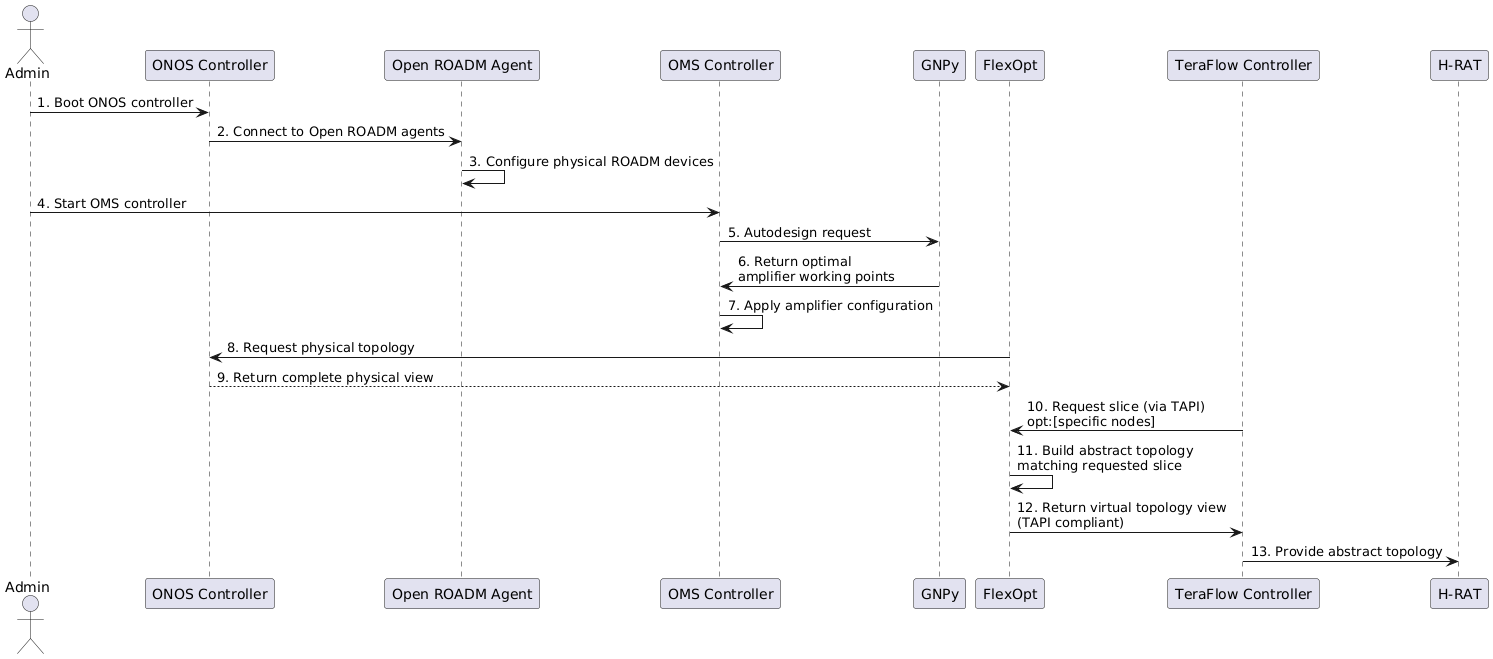}
    \caption{Workflow for dynamic network discovery, optimal configuration, and slice-specific topology abstraction enabling resource allocation and service orchestration over the optical infrastructure.}
    \label{fig:discovery}
\end{figure*}
This paper presents a demonstration of a fully functional OSaaS platform that supports deploying 3GPP network slices on top of a per-tenant optical network slices instantiated over a disaggregated testbed. 
The proposed architecture combines several key components: the ONOS SDN controller manages the optical data plane and media channel configuration; the FlexOpt optical orchestrator provides optical topology abstraction, differentiated network views with resource isolation and exposes TAPI-compliant NBI interfaces, both in terms of optical network slicing as well as in terms of service provisioning within a given tenant/slice; the TeraFlow SDN controller handles slice management and IP-over-WDM (IPoWDM) control; and H-RAT, a holistic resource allocation tool, computes optimal paths for service provisioning. A service orchestrator triggers the end-to-end workflow, enabling dynamic provisioning of tenant-specific optical slices.
This demonstration showcases the feasibility of OSaaS through a disaggregated and multi-layer control stack, highlighting the interaction between resource abstraction, automation, and standardized interoperability. The work paves the way for future deployments of programmable and tenant-aware optical infrastructures in both research and production networks.
\begin{figure*}[t!]
    \centering
    \includegraphics[width=1\textwidth]{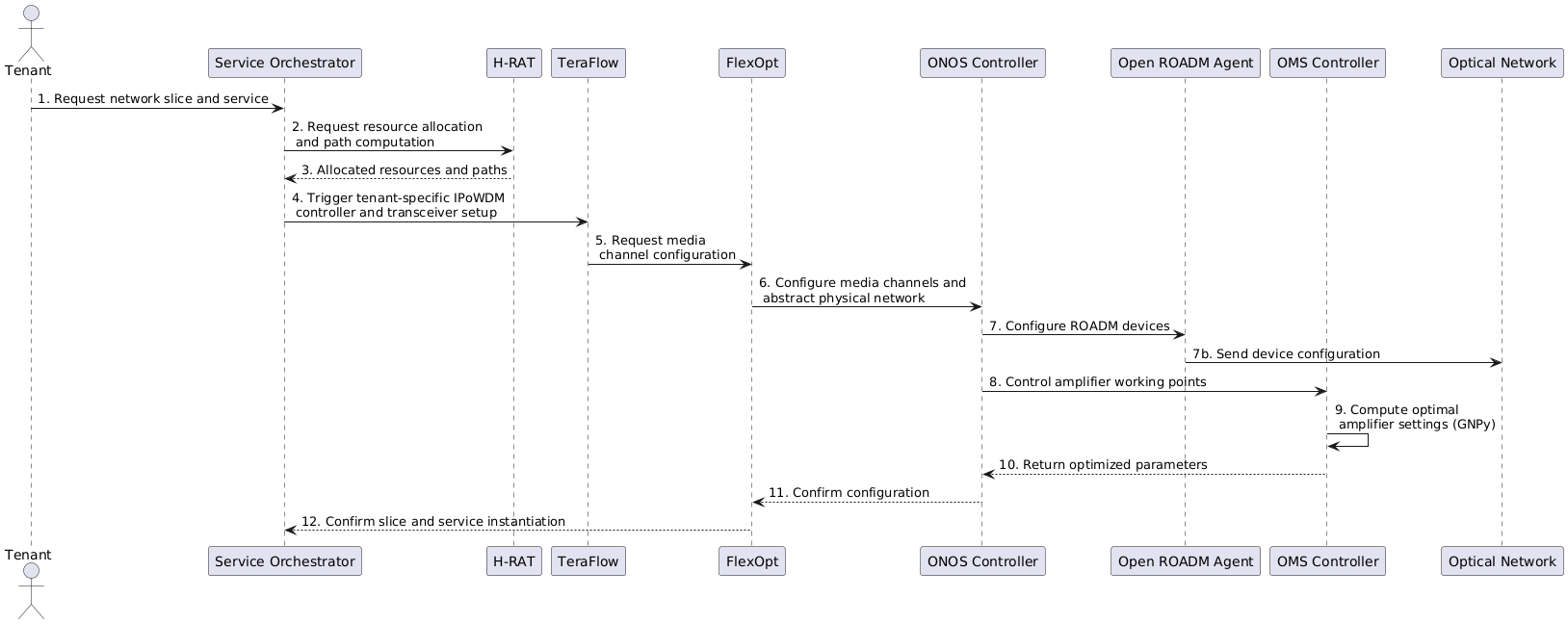}
    \caption{Sequence diagram depicting the end-to-end workflow for network slice request, resource allocation, and optical network configuration, from tenant interaction to final device setup.}
    \label{fig:sequence-chart}
\end{figure*}
\section{Architecture \& technologies}
This demo operates in the context of disaggregated optical networks, where slicing and service differentiation are essential to support OSaaS use cases. The optical infrastructure used in this demo is based on a core optical network testbed hosted at Links Foundation labs in Turin and is detailed in Fig.~\ref{fig:topology-control}.
The physical topology consists of three ROADM nodes interconnected by three Optical Multiplexing Sections (OMSs). Each OMS is composed of fiber spans and EDFA amplifiers, simulating realistic long-haul transport segments.
\todo[inline]{ALESSIO}
The optical nodes are controlled using ONOS SDN controller~\cite{giorgetti2020control}, which manages the configuration of media channels. 
Each ROADM in the setup has been equipped with an OpenROADM-compliant agent that exposes a standard northbound interface (NBI) while internally translating requests into the vendor-specific southbound interfaces (SBI). 
Furthermore, the ONOS controller has been extended to support an OpenROADM-specific driver, enabling it to interact with the ROADM agents using the OpenROADM model~\cite{schips2024enhancing}.
The EDFA amplifiers, due to the multi-vendor and disaggregated nature of the setup, are controlled via a custom software stack that has been validated in previous research~\cite{ambrosone2023open}. This is necessary as the OMSs are built using devices from different vendors, both across OMSs and with respect to the ROADMs. During the demo, the entire setup can be remotely controlled, and media channel provisioning is visualized through the ONOS GUI.
All the EDFAs configuration and data retrieve are performed using OpenConfig-compliant NBI interface.~\cite{riccardoICTON}.
This enables vendor-neutral monitoring and configuration of key optical parameters such as amplifier gain and tilt. To determine the optimal working points for each OMS, the controller leverages GNPy as a digital twin of the physical network. GNPy simulates the optical layer with high fidelity and provides guidance on optimal amplifier settings that maximize end-to-end Quality of Transmission (QoT) while ensuring safe operational margins. In this context, GNPy is not only a validation tool but also a central component for closed-loop optimization. This architectural approach is in line with the Open Line Controller paradigm proposed in~\cite{ambrosone2025ondm}, where model-driven control and telemetry-driven optimization are key enablers for managing partially disaggregated optical networks.
On top of ONOS, the FlexOpt~\cite{casellas2024photonic} orchestrator is deployed. FlexOpt allows the dynamic instantiation/creation of optical line system network slices, which correspond to specific partition of the line system in terms of resources such as nodes, links, ports or spectrum ranges \cite{, LeisbethTapi} which, in turn, may show shared or exclusive ownership across the different tenants. As stated, FlexOpt exposes a TAPI based NBI and, within the context of a specific tenant/view, it enables the creation of media channel services (optical channels) restricted to a specific set of resources. To achieve this, it interacts with ONOS via its REST APIs to first retrieve the underlying physical infrastructure and later on, upon request, configure the underlying optical resources to provision optical connections. 
\todo[inline]{--Telefonica --}
Above this abstraction, resource management and slice orchestration are handled by the TeraFlow SDN controller~\cite{gifre2022experimental}. TeraFlow is responsible not only for tenant-specific resource isolation but also for managing IPoWDM scenarios, including transceiver control such as power and wavelength adjustment. Each tenant is assigned a dedicated IP controller that orchestrates the end-to-end connection establishment using the virtualized optical layer provided by FlexOpt.
\todo[inline]{--e-lighthouse-- }
The remaining two components complete the control stack:
H-RAT (Holistic Resource Allocation Tool). This module is in charge of selecting the optimal resource allocation and path computation across the available slices.
Service Orchestrator: This module is responsible for triggering the provisioning workflow, interacting with H-RAT, FlexOpt, and TeraFlow to instantiate tenant services.

\section{Implementation \& Demo Workflow}

\textit{Phase I:} Network Discovery (shown in Fig.\ref{fig:discovery})\\\noindent
\textit{1.} At boot, the ONOS controller establishes connections with OpenROADM agents on each device.\\\noindent
\textit{2.} Concurrently, the OMS controller runs a local GNPy instance to compute and configure optimal amplifier working points.\\\noindent
\textit{3.} After device discovery and configuration, FlexOpt queries ONOS via REST APIs to obtain the full physical topology.\\\noindent
\textit{4.} TeraFlow request to FlexOpt tenant-specific abstract topologies (network slices) through a TAPI interface, allowing slice instantiation with defined nodes and links.\\\noindent
\textit{5.} TeraFlow interacts with H-RAT to evaluate slices and compute feasible optical paths.\\\noindent
\textit{6.} These paths are forwarded to the Service Orchestrator to prepare for service deployment.\\\noindent\\\noindent
\textit{Phase II:} Connection Setup and Configuration (shown in Fig.\ref{fig:sequence-chart})\\\noindent
\textit{1.} A tenant requests a network slice and service via the Service Orchestrator.\\\noindent
\textit{2.} The Orchestrator asks the Slice Manager (with H-RAT) for resource allocation and path selection.\\\noindent
\textit{3.} Resources are allocated and optimal paths selected.\\\noindent
\textit{4.} The Service Orchestrator triggers the tenant-specific IPoWDM controller within TeraFlow to configure transceivers and IP-layer settings.\\\noindent
\textit{5.} TeraFlow requests media channel setup from FlexOpt, which uses ONOS REST APIs to configure media channels and manage the virtualized physical network.\\\noindent
\textit{6.} ONOS programs ROADM devices via OpenROADM agents through southbound interfaces.\\\noindent
\textit{7.} Finally, confirmation is sent back to the Service Orchestrator.

\section{Demo Innovation and Audience Relevance}
\todo[inline]{pablo}
The core innovation of this demo lies in the instantiation of 3GPP network slices, as sets of interconnected functions tailored to specific customer needs on top of optical transport network slices, which are dynamic partitions of the optical infrastructure managed by individual tenants. The architecture leverages a real photonic mesh network and integrates several functional components to perform slice computation, function placement, and resource allocation.
What makes this demo particularly appealing to conference attendees is the ability to observe a fully remote-controlled setup in action. The system's behavior is showcased both through the GUIs of the various orchestration and control components and by displaying real-time telemetry data collected from the network devices. During the demonstration, different topology and frequency slices will be instantiated, offering a comprehensive view of how end-to-end slicing can be achieved across layers.

\clearpage
\section{Acknowledgements}
This work has been supported by the EU Horizon Europe research and innovation program, ALLEGRO Project, GA No. 101092766.


\balance
\printbibliography


\end{document}